# Turbulent broadening of electron heat-flux width in electromagnetic gyrokinetic simulations of a helical scrape-off layer model

N. R. Mandell, G. W. Hammett, A. Hakim, et al.

### COLLECTIONS

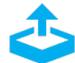

### ARTICLES YOU MAY BE INTERESTED IN

Kinetic Landau-fluid closures of non-Maxwellian distributions
Physics of Plasmas **29**, 042116 (2022); https://doi.org/10.1063/5.0083108

A survey of pedestal magnetic fluctuations using gyrokinetics and a global reduced model for microtearing stability
Physics of Plasmas **29**, 042503 (2022); https://doi.org/10.1063/5.0084842

Collective plasma effects of electron–positron pairs in beam-driven QED cascades
Physics of Plasmas **29**, 042117 (2022); https://doi.org/10.1063/5.0078969

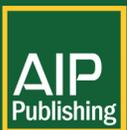

# Turbulent broadening of electron heat-flux width in electromagnetic gyrokinetic simulations of a helical scrape-off layer model

N. R. Mandell,[1,2,a)] G. W. Hammett,[3] A. Hakim,[3] and M. Francisquez[3]

AFFILIATIONS

[1]Department of Astrophysical Sciences, Princeton University, Princeton, New Jersey 08543, USA
[2]MIT Plasma Science and Fusion Center, Cambridge, Massachusetts 02139, USA
[3]Princeton Plasma Physics Laboratory, Princeton, New Jersey 08543, USA

Note: This paper is part of the Special Collection: Papers from the Sherwood 2021 Fusion Theory Conference.
[a)]Author to whom correspondence should be addressed: nrm@mit.edu

## ABSTRACT

We demonstrate that cross field transport in the scrape-off layer (SOL) can be moderately increased by electromagnetic effects in high-beta regimes, resulting in broadening of the electron heat-flux width on the endplates. This conclusion is taken from full-$f$ electromagnetic gyrokinetic simulations of a helical SOL model that roughly approximates the SOL of the National Spherical Torus Experiment. The simulations have been performed with the G<small>KEYLL</small> code, which recently became the first code to demonstrate the capability to simulate electromagnetic gyrokinetic turbulence on open magnetic field lines with sheath boundary conditions. We scan the source rate and thus $\beta$, so that the normalized pressure gradient (the MHD ballooning parameter $\alpha \propto \partial \beta / \partial r \propto \beta / L_p$) is scanned over an experimentally relevant range, $\alpha = 0.3 - 1.5$. While there is little change in the pressure gradient scale length $L_p$ near the midplane as beta is increased, a 10% increase in cross field transport near the midplane results in an increase in the electron heat-flux width $\lambda_q$ and a 25% reduction of the peak electron heat flux to the endplates.

## I. INTRODUCTION

One of the primary challenges for burning-plasma fusion devices is the power exhaust problem: how to mitigate the extreme heat fluxes to the materials surrounding plasma. Heat from the core is transported across the last-closed-flux-surface and exhausted in the scrape-off layer (SOL), where the heat quickly flows along open field lines to the material walls of the device in the divertor. We must ensure the heat load from devices, such as ITER, can be reduced below material limits in order to avoid damage to the divertor plates and the introduction of impurities that degrade fusion performance. Importantly, mitigation strategies for the large heat loads must be careful to avoid degrading the high temperatures in the core required to sustain a burning plasma.[1]

A key parameter is the width of the heat flux channel, $\lambda_q$, since spreading the heat over a larger area reduces the peak heat load. The heat-flux width is determined by competition between parallel transport along the background magnetic field and cross field turbulent transport.[2,3] An empirical scaling computed from a multi-machine database has shown that $\lambda_q$ varies strongest with the inverse of the poloidal magnetic field strength.[4,5] The validity of extrapolating this empirical scaling to future devices, such as ITER, is an important issue that can be addressed by first-principles modeling. For example, XGC1 electrostatic gyrokinetic simulations reproduce the empirical scaling for existing tokamaks but predict $\lambda_q$ that is about six times wider when extrapolated to ITER, with the widening due to an increase in trapped-electron turbulence.[6,7]

Thus, insights from theory and numerical modeling of the plasma boundary are critical to solving the challenging power-exhaust issue. Significant progress has been made in modeling the edge and SOL using fluid[8–14] and gyrokinetic[15–22] models. In this work, we investigate the effects of magnetic fluctuations on SOL turbulence and the resulting width of the heat flux channel. The edge/SOL region features steep pressure gradients, especially in the H-mode transport barrier and SOL regions, which can contribute to the importance of electromagnetic effects. Experimental evidence has indicated that the edge plasma state is controlled by electromagnetic drift-interchange dynamics.[23–25] In this regime, the parallel electron dynamics is no longer very fast relative to the drift turbulence, so electrons can no longer be treated adiabatically.[26] In some cases, this can lead to coupling of the perpendicular vortex motions and kinetic shear Alfvén waves, which results in field-line bending.[27] Electromagnetic and sheath effects on scrape-off layer dynamics have been investigated using drift-Braginskii fluid models, as considered in some of the fluid studies mentioned above. Plasma resistivity and sheath resistivity have been identified as mechanisms that can destabilize interchange and resistive ballooning modes below the ideal limit.[28–31] Electromagnetic modifications to blob dynamics have also been investigated, showing that a finite Alfvén speed can allow electrical disconnection from the walls and produce faster blob velocities.[32–34]

The GKEYLL code has recently become the first code to demonstrate the capability of simulating electromagnetic gyrokinetic (EMGK) turbulence on open magnetic field lines with sheath boundary conditions.[20,35,36] Previous GKEYLL results include electrostatic studies of the SOL of the National Spherical Torus Experiment (NSTX),[19] the Texas Helimak,[37] and the Large Plasma Device (LAPD).[38] While we have been focused on demonstrating the capability of simulating the SOL region, GKEYLL will eventually be used to simulate the pedestal region as well, where it is known that kinetic ballooning and peeling–ballooning modes play an important role in setting the height and width of pedestals,[39,40] and microtearing modes are also important in some regimes.[41,42] Thus, it is important to demonstrate a robust capability of handling electromagnetic fluctuations in an edge gyrokinetic code.

In this work, we use GKEYLL to perform full-$f$ electromagnetic gyrokinetic simulations of a helical scrape-off layer (SOL) model that roughly approximates the SOL of NSTX. The remainder of this paper is organized as follows: In Sec. II, we describe the electromagnetic gyrokinetic model used for the simulations. We describe the simulation geometry and setup in Sec. III. Section IV presents the primary results, including the findings that electromagnetic effects can broaden the electron heat flux width in high beta cases due to increased cross field transport, despite the fact that midplane profiles are relatively unmodified. Conclusions are given in Sec. V.

## II. ELECTROMAGNETIC GYROKINETIC MODEL

We model turbulence by solving the full-$f$ electromagnetic gyrokinetic (EMGK) system in the long-wavelength (drift-kinetic) limit. The electromagnetic fluctuations are treated via the symplectic formulation,[43] so that the parallel velocity $v_\parallel$ is used as a coordinate. [This is different than the "Hamiltonian" formulation commonly used in particle-in-cell (PIC) approaches that use the canonical momentum $p_\parallel$ as a coordinate, though both approaches can be expressed in terms of Poisson brackets and analytically conserve the same conservation laws exactly, such as energy conservation.]

In the long-wavelength limit, the gyrokinetic equation describes the evolution of the guiding-center distribution function $f_s = f_s(\mathbf{R}, v_\parallel, \mu; t)$ for species $s$, where $\mathbf{R} = (x, y, z)$ is the guiding-center position, $v_\parallel$ is the parallel velocity, and $\mu = m_s v_\perp^2/(2B)$ is the magnetic moment. In conservative form, we have

$$\frac{\partial(\mathcal{J}f_s)}{\partial t} + \nabla \cdot (\mathcal{J}\dot{\mathbf{R}}f_s) + \frac{\partial}{\partial v_\parallel}\left(\mathcal{J}\dot{v}_\parallel^H f_s\right) - \frac{\partial}{\partial v_\parallel}\left(\mathcal{J}\frac{q_s}{m_s}\frac{\partial A_\parallel}{\partial t}f_s\right) = \mathcal{J}C[f_s] + \mathcal{J}S_s, \quad (1)$$

where the nonlinear phase-space characteristics are given by

$$\dot{\mathbf{R}} = \frac{\mathbf{B}^*}{B_\parallel^*}v_\parallel + \frac{\hat{\mathbf{b}}}{qB_\parallel^*} \times (\mu \nabla B + q\nabla\Phi), \quad (2)$$

$$\dot{v}_\parallel = \dot{v}_\parallel^H - \frac{q}{m}\frac{\partial A_\parallel}{\partial t} = -\frac{\mathbf{B}^*}{mB_\parallel^*} \cdot (\mu \nabla B + q\nabla\Phi) - \frac{q}{m}\frac{\partial A_\parallel}{\partial t}, \quad (3)$$

with $\Phi$ being the electrostatic potential and $A_\parallel$ the parallel magnetic vector potential. Collisions $C[f_s]$ and sources $S_s$ are included on the right-hand side of (1) with the specific forms of these terms as used in this work detailed below. Here, $B_\parallel^* = \hat{\mathbf{b}} \cdot \mathbf{B}^*$ is the parallel component of the effective magnetic field $\mathbf{B}^* = \mathbf{B} + (m_s v_\parallel/q_s)\nabla \times \hat{\mathbf{b}} + \delta\mathbf{B}$, where $\mathbf{B} = B\hat{\mathbf{b}}$ is the equilibrium magnetic field and $\delta\mathbf{B} = \nabla \times (A_\parallel \hat{\mathbf{b}})$ is the perturbed magnetic field, neglecting compressional magnetic fluctuations. The Jacobian of the gyrocenter coordinates is $\mathcal{J} = B_\parallel^*/m_s$, and we make the approximation $\hat{\mathbf{b}} \cdot \nabla \times \hat{\mathbf{b}} \approx 0$ so that $B_\parallel^* \approx B$. The species charge and mass are $q_s$ and $m_s$, respectively. In ((3)), note that we have separated $\dot{v}_\parallel$ into a term that comes from the Hamiltonian, $\dot{v}_\parallel^H$, and another term proportional to the inductive component of the parallel electric field, $\partial A_\parallel/\partial t$. We use this notation for convenience, so that the time derivative of the parallel vector potential $A_\parallel$ appears explicitly, which is characteristic of the symplectic formulation of EMGK.

The electrostatic potential is determined by the quasi-neutrality condition in the long-wavelength limit, which takes the following form of the Poisson equation:

$$-\nabla \cdot (\epsilon_\perp \nabla_\perp \Phi) = \sum_s q_s \int f_s \, d^3\mathbf{v}, \quad (4)$$

with $d^3\mathbf{v} = 2\pi dv_\parallel d\mu \mathcal{J}$ and

$$\epsilon_\perp = \sum_s \frac{m_s n_{0s}}{B^2}. \quad (5)$$

Here, we use a linearized polarization density $n_0$ that we take to be a constant in time, which is consistent with neglecting a second-order $E \times B$ energy term in the Hamiltonian. While the validity of this Boussinesq-type approximation in the SOL can be questioned due to large density fluctuations (and we plan to eventually improve on this approximation), a linearized polarization density is commonly used for computational efficiency.[19,44] The magnetic vector potential is determined by the parallel Ampère equation,

$$-\nabla_\perp^2 A_\parallel = \mu_0 \sum_s q_s \int v_\parallel f_s \, d^3\mathbf{v}. \quad (6)$$

Note that we can also take the time derivative of this equation to get a generalized Ohm's law, which can be solved directly for $\partial A_\parallel/\partial t$, the inductive component of the parallel electric field $E_\parallel$,[45–47]

$$-\nabla_\perp^2 \frac{\partial A_\parallel}{\partial t} = \mu_0 \sum_s q_s \int v_\parallel \frac{\partial f_s}{\partial t} d^3\mathbf{v}. \quad (7)$$

Writing the gyrokinetic equation as

$$\frac{\partial(\mathcal{J}f_s)}{\partial t} = \frac{\partial(\mathcal{J}f_s)^\star}{\partial t} + \frac{\partial}{\partial v_\parallel}\left(\mathcal{J}\frac{q_s}{m_s}\frac{\partial A_\parallel}{\partial t}f_s\right), \quad (8)$$

where $\partial(\mathcal{J}f_s)^\star/\partial t$ denotes all the terms in the gyrokinetic equation (including sources and collisions) except the $\partial A_\parallel/\partial t$ term, Ohm's law can be rewritten (after an integration by parts) as

$$\left(-\nabla_\perp^2 + \sum_s \frac{\mu_0 q_s^2}{m_s}\int f_s\, d^3\boldsymbol{v}\right)\frac{\partial A_\parallel}{\partial t} = \mu_0 \sum_s q_s \int v_\parallel \frac{\partial f_s^\star}{\partial t}\, d^3\boldsymbol{v}. \quad (9)$$

In the GKEYLL code, we use (9) to compute $\partial A_\parallel/\partial t$ directly and use this to evolve $A_\parallel$ in time with (6) only used as an initial condition (see Ref. 20 for more details).

To model the effect of collisions, we use a conservative Lenard–Bernstein (or Dougherty) collision operator,[48,49]

$$\mathcal{J}C[f_s] = \sum_r \nu_{sr}\left\{\frac{\partial}{\partial v_\parallel}\left[(v_\parallel - u_{\parallel sr})\mathcal{J}f_s + v_{tsr}^2 \frac{\partial(\mathcal{J}f_s)}{\partial v_\parallel}\right] + \frac{\partial}{\partial \mu}\left[2\mu\mathcal{J}f_s + 2\mu\frac{m_s}{B}v_{tsr}^2\frac{\partial(\mathcal{J}f_s)}{\partial \mu}\right]\right\}, \quad (10)$$

where like-species collisions use $u_{\parallel sr} = u_{\parallel s}$, $v_{tsr} = v_{ts}$, and these quantities are given by

$$n_s u_{\parallel s} = \int v_\parallel f_s\, d^3\boldsymbol{v}, \quad (11)$$

$$n_s u_{\parallel s}^2 + 3n v_{ts}^2 = \int\left(v_\parallel^2 + 2\mu B/m_s\right)f_s\, d^3\boldsymbol{v}, \quad (12)$$

with $n_s = \int f_s\, d^3\boldsymbol{v}$ (see Ref. 50 for more details). Cross-species collisions among electrons and ions are also modeled.[51] This collision operator contains the effects of drag and pitch-angle scattering, and it conserves number, momentum, and energy density. Consistent with our present long-wavelength treatment of the gyrokinetic system, finite-Larmor-radius effects are ignored. Note that in this model collision operator, the collision frequency $\nu$ is velocity-independent, i.e., $\nu \neq \nu(v)$.

## III. SIMULATION SETUP

As a step toward modeling the tokamak scrape-off layer, we consider a simple helical scrape-off layer model. In this configuration, the magnetic field is composed of a toroidal component $B_\varphi$ and a vertical component $B_v$, giving helical field lines. All field lines are open, terminating on material walls at the top and bottom of the device. This configuration is also known as a simple magnetized torus (SMT) and has been experimentally studied via devices such as the Helimak[52] and TORPEX.[53] Despite the relative simplicity of the helical SMT configuration, it contains unfavorable magnetic curvature in the presence of density and pressure gradients. This produces the interchange instability that drives turbulence and blob dynamics in the SOL. We use parameters roughly modeling the SOL of the National Spherical Torus Experiment (NSTX) at PPPL. Note that the helical geometry has no favorable curvature region. Thus, one can think of our geometry as a rough approximation of the SOL of a double-null diverted configuration, since the outboard SOL flux surfaces in this configuration do not sample the good curvature region of the tokamak. However, we have not included magnetic shear in our simulation geometry, which, in reality, would be quite strong near the X-points of a double-null configuration, including magnetic shear and X-points that are left for future work.

We simulate a flux-tube-like domain[54] that helically wraps around the torus and terminates on conducting plates at each end. For this, we use a field-aligned coordinate system,[55] with $x$ being the radial coordinate, $z$ the coordinate along the field lines, and $y$ the binormal coordinate that labels field lines at constant $x$ and $z$. One can think of these coordinates roughly mapping to physical cylindrical coordinates $(R, \varphi, Z)$ via $R = x$, $\varphi = y/(R_c \sin\chi) + z\cos\chi/x$, $Z = z\sin\chi$. We take the field-line pitch angle $\chi = \sin^{-1}(B_v/B)$ to be constant, with $B_v$ being the vertical component of the magnetic field (analogous to the poloidal field in typical tokamak geometry) and $B$ the total magnitude of the background magnetic field. Furthermore, $R_c = R_0 + a$ is the radius of curvature at the center of the simulation domain, with $R_0$ being the device major radius and $a$ the minor radius. We neglect all geometrical factors arising from the non-orthogonal coordinate system, except for the assumption that perpendicular gradients of $f$ are much stronger than parallel gradients. Thus, we can approximate

$$(\nabla \times \hat{\mathbf{b}}) \cdot \nabla f(x,y,z) \approx \left[(\nabla \times \hat{\mathbf{b}})\cdot\nabla y\right]\frac{\partial f}{\partial y} = -\frac{1}{x}\frac{\partial f}{\partial y}, \quad (13)$$

where we have used $\boldsymbol{B} \approx B_{\text{axis}}(R_0/x)\boldsymbol{e}_z$, with $B_{\text{axis}}$ being the magnetic field strength at the magnetic axis, and neglected the contribution of the small vertical field $B_v$. This means that the magnetic (curvature plus $\nabla B$) drift,

$$\boldsymbol{v}_d = \frac{mv_\parallel^2}{qB}\nabla \times \hat{\mathbf{b}} + \frac{\mu}{qB}\hat{\mathbf{b}} \times \nabla B, \quad (14)$$

is purely in the $y$ direction,

$$\boldsymbol{v}_d \cdot \nabla y = -\left(\frac{mv_\parallel^2 + \mu B}{qB}\right)\frac{1}{x} = -\frac{mv_\parallel^2 + \mu B}{qB_{\text{axis}}R_0}, \quad (15)$$

with $\boldsymbol{v}_d \cdot \nabla x = \boldsymbol{v}_d \cdot \nabla z = 0$. Thus, this simplified geometry has constant magnetic curvature (the curvature does not vary along the field line, so there is no good curvature region to produce conventional ballooning-mode structure), and we have neglected magnetic shear in the present setup. As shown in Appendix 5.B of Ref. 37, these approximations are consistent with taking the limit $B_v \ll B$, which results in a purely toroidal field. The GKEYLL code has also been generalized to include general toroidal geometry with all geometrical factors accounted for;[36] these results will be reported elsewhere.

Taking NSTX-like parameters, we use $R_0 = 0.85$ m, $a = 0.5$ m, and $B_{\text{axis}} = 0.5$ T. The simulation box is centered at $(x_0, y_0, z_0) = (1.34\,\text{m}, 0, 0)$ with dimensions $L_x = 56\rho_{s0} \approx 16.6$ cm, $L_y = 100\rho_{s0} \approx 29.1$ cm, and $L_z = L_{\text{pol}}/\sin\chi = 8$ m, where $L_{\text{pol}} = 2.4$ m and $\rho_{s0} = c_{s0}/\Omega_i$ for typical reference temperatures $T_0 \sim 40$ eV. The radial boundary conditions model conducting walls at the radial ends of the domain, given by the Dirichlet boundary condition $\Phi = A_\parallel = 0$. The condition $\Phi = 0$ prevents $E \times B$ flows into walls, while $A_\parallel = 0$ makes it so that (perturbed) field lines never intersect the walls. For the latter, one can think of image currents in the conducting wall that mirror currents in the domain, resulting in exact cancelation of the perpendicular magnetic fluctuations at the wall. Also, note that in this simple magnetic geometry, the magnetic drifts do not have a radial component. Thus, these radial boundary conditions on the fields are sufficient to ensure that there is no flux of the distribution function to the radial boundaries. These radial boundary conditions are a simplifying approximation, particularly at the inner boundary where the physical inner boundary for the SOL should be the interface with the closed-field-line region of the tokamak. Since we do not presently model the closed-field-line region (this will be included in future work), we have taken these simple Dirichlet boundary conditions and have ensured that the radial boundaries of the simulation domain are far enough away so as to not affect the results in the interior region of interest. Periodic boundary conditions are used in the $y$ direction. Conducting-sheath boundary conditions are applied to the distribution function in the $z$ direction, which model the Debye sheath (the dynamics of which is beyond the gyrokinetic ordering) by reflecting low-energy electrons and absorbing high energy electrons and all ions. This involves solving the gyrokinetic Poisson equation to evaluate the potential at the $z$ boundary, corresponding to the sheath entrance, and using the resulting sheath potential to determine a cutoff velocity below which electrons are reflected by the sheath.[19,38] This boundary condition allows local current fluctuations in and out of the sheath, unlike the standard logical sheath boundary condition[56] that imposes that the ion and electron currents at the sheath entrance are equal at all times. The fields do not require a boundary condition in the $z$ direction since only perpendicular derivatives appear in the field equations. The velocity-space grid has extents $-4v_{ts} \leq v_\parallel \leq 4v_{ts}$ and $0 \leq \mu \leq 6T_0/B_0$, where $v_{ts} = \sqrt{T_0/m_s}$ and $B_0 = B_{axis}R_0/R_c$.

To model particles and heat from the core crossing the separatrix and entering the SOL, we include an *ad hoc* non-drifting Maxwellian source of ions and electrons,

$$S_{i,e} = \frac{n_S(x,z)}{(2\pi T_S/m_{i,e})^{3/2}} \exp\left(-\frac{m_{i,e}v^2}{2T_S}\right), \quad (16)$$

with source temperature $T_S = 70$ eV for both species. The source density is given by

$$n_S(x,z) = \begin{cases} S_0 \exp\left(\frac{-(x-x_S)^2}{(2\lambda_S)^2}\right), & |z| < L_z/4, \\ 0, & \text{otherwise,} \end{cases} \quad (17)$$

so that the source is localized in the region $x_S - 3\lambda_S < x < x_S + 3\lambda_S$, and we take $x_S = 1.3$ m and $\lambda_S = 0.005$ m. The localization of the source near the midplane in $z$ models ballooning-like transport crossing the separatrix from the core near the outboard midplane of the tokamak. The source particle rate $S_0$ is chosen so that the total (ion plus electron) source power matches the desired power into the simulation domain, $P_{src}$. Since we only simulate a flux-tube-like fraction of the whole SOL domain, $P_{src}$ is related to the total SOL power, $P_{SOL}$, by $P_{src} = P_{SOL}L_yL_z/(2\pi R_cL_{pol}) \approx 0.115 P_{SOL}$. The simulations reach a quasi-steady state with the sources balanced by end losses to the sheath, after which time-average quantities can be computed. Note that in experiments, neutral recycling near the endplates can be the dominant particle source (if there is no gas puffing); we do not model neutrals in this work with a kinetic neutral model in progress.[57,58] Figure 1 shows the source particle rate profile as a function of the radial ($x$) and along-the-field-line ($z$) coordinates along with the boundary conditions. The black dotted lines denote the full width at half maximum (FWHM) of the source profile. The initial density and temperature profiles are taken to be proportional to the source profiles with the exact form chosen as in Ref. 19 from the steady-state solution of one-dimensional fluid equations.

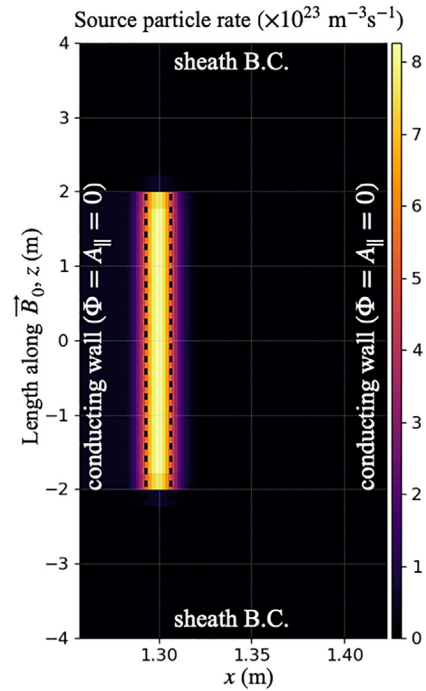

FIG. 1. Since we do not include a closed-field-line region in our simulations, we use a source localized near the inner radial edge of the domain to model particles and heat crossing the separatrix and entering the scrape-off layer. This diagram shows the source particle rate profile as a function of the radial ($x$) and along-the-field-line ($z$) coordinates. The temperature of all source particles is $T_S = 70$ eV for both electrons and ions. Black dotted lines denote the radial full width at half maximum (FWHM) of the profile, which extends from $-2 < z < 2$ m. Also shown are the boundary conditions: Dirichlet at the radial boundaries with $\Phi = A_\parallel = 0$ and sheath boundary conditions along the field line. Not pictured are periodic boundary conditions in the $y$ direction (out-of-plane here).

The simulations in this work were performed with the EMGK module of the GKEYLL plasma simulation framework. This module employs a discontinuous Galerkin (DG) discretization scheme for the EMGK system that conserves energy (in the continuous-time limit) and avoids the Ampère cancelation problem.[20] Our simulations use piecewise-linear ($p=1$) basis functions with $(N_x, N_y, N_z, N_{v_\parallel}, N_\mu) = (48, 96, 18, 10, 5)$ being the number of cells in each dimension. Note that for $p=1$ DG, there are two degrees of freedom per dimension in each cell, so one should double each of these numbers to obtain the equivalent number of degrees of freedom for comparison with other gyrokinetic codes. We also artificially reduce the collision frequency to 10% of its physical value to avoid an expensive time step reduction from large collisionality (this could be avoided in the future by using an implicit discretization of the collision operator). This also allows us to isolate electromagnetic effects that change with density from collisional effects that also scale with density.[59] In reality, collisional viscosity and magnetic induction compete to slow parallel electron dynamics, with the slowest timescale dominating the behavior.[26]

We perform a parameter scan of the source particle rate, which roughly controls the density in our flux-driven simulations. The base case is based on the nominal experimental heating power for typical H-mode cases on NSTX,[60] $P_{\text{SOL,base}} = 5.4$ MW. We then scan the source particle rate by taking $P_{\text{SOL}} = \hat{n} P_{\text{SOL,base}}$, with $\hat{n} = \{1, 2, 3.5, 5\}$ at constant source temperature ($T_s = 70$ eV). As a result, the scaling factor $\hat{n}$ is roughly proportional to the density and $\beta$. Electromagnetic fluctuations are included in all of the simulations presented here. A corresponding $\hat{n} = 1$ electrostatic case has also been studied, and the results are nearly identical to the $\hat{n} = 1$ electromagnetic case. Thus, we will not include the electrostatic case in the following analysis, and instead simply use the $\hat{n} = 1$ case as a proxy for the electrostatic limit.

Snapshots of the $\hat{n} = 1$ (left) and $\hat{n} = 5$ (right) cases are shown in Fig. 2. The snapshots are taken in the perpendicular plane at the midplane ($z = 0$). The electron density is shown in the top row, while the electrostatic potential is shown in the bottom row, with gray lines denoting constant potential surfaces. The density structures are similar in both cases, with blobs that propagate radially outwards. In the lower $\beta$ cases, there is a tendency toward monopole potential structures and adiabatic dynamics, which causes the blobs to spin due to the Boltzmann spinning effect.[61] The presence of strong adiabatic dynamics in the electrostatic limit is a feature of the plasma being highly conducting, which is likely influenced by our choice to scale down the collisionality for computational reasons.

## IV. RESULTS: MIDPLANE PROFILES, HEAT-FLUX WIDTH, AND TRANSPORT

In Fig. 3, we compute time- and $y$-averaged midplane ($z = 0$) profiles of density, temperature, and $\beta$ for each of the $\hat{n} = \{1, 2, 3.5, 5\}$ cases. All time averages (here and in later plots) are taken over a period $\sim 500 - 700 \, \mu$s after the profiles have reached steady state. Note that the physically relevant region of these midplane profiles extends from the source peak at $x = 1.3$ m to roughly $x = 1.4$ m, where boundary approximations do not have a significant effect on the profiles. Ion guiding-center quantities are shown on the left, while electron quantities are shown on the right. The density and $\beta$ plots have been normalized to the source particle rate scaling factor $\hat{n}$, so that the shape and gradient scale lengths of the profiles can be easily compared as $\beta \sim \hat{n}$ increases. The density (and $\beta$) scales with $\hat{n}$ while the temperature does not, as one would expect when increasing the source power at fixed source temperature. Apart from this, there does not appear to be significant change in the gradient scale lengths as $\beta$ increases. There is a slight drop in the peak (normalized) density in the $\hat{n}=5$, but the density gradient scale length outside of the source region (shaded) is nearly the same for all cases. This is somewhat of a null result, indicating that electromagnetic effects are not influencing the midplane gradient scale lengths. An interesting feature is that we observe $T_i/T_e \sim 2$, which is consistent with experimental measurements indicating $1 \lesssim T_i/T_e \lesssim 4$ in the SOL.[62] In a sheath-limited SOL, low-energy electrons are retained by the sheath boundary effects while high-energy energy electrons are lost, so that more heat is lost to the sheath by electrons than by ions.

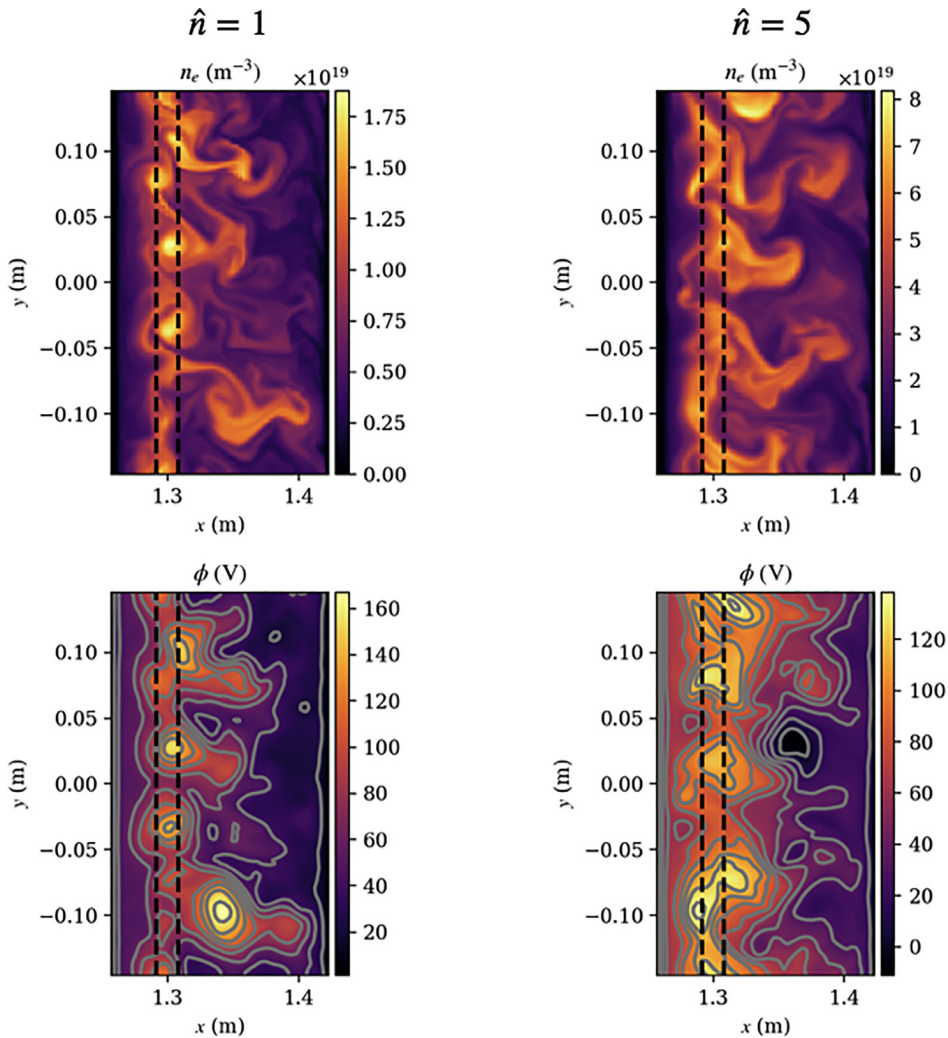

**FIG. 2.** Snapshots from the $\hat{n} = 1$ (left) and $\hat{n} = 5$ (right) cases, taken in the perpendicular plane at the midplane ($z = 0$) at $t = 540 \, \mu$s. The electron density is shown in the top row, and the bottom row shows the electrostatic potential with gray lines denoting constant potential contours. Black dotted lines denote the FWHM of the source profile.

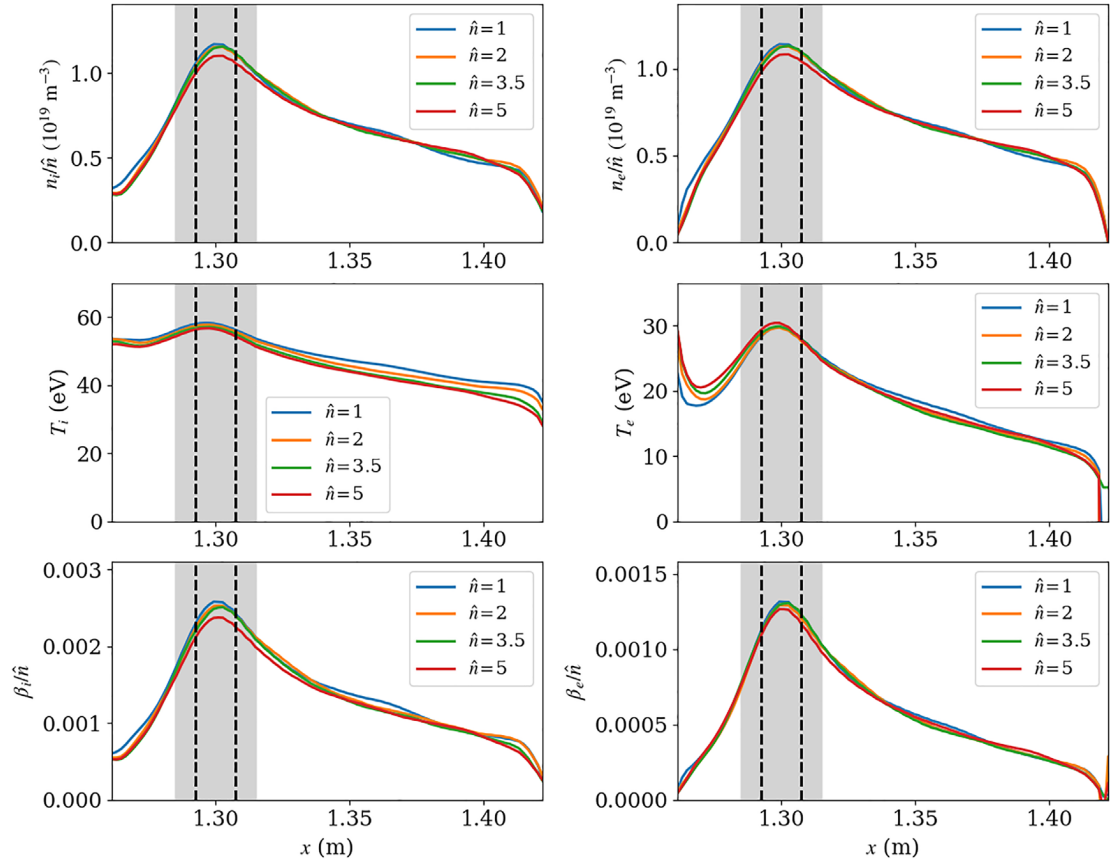

**FIG. 3.** Time- and $y$-averaged midplane ($z=0$) profiles of density, temperature, and $\beta$ for each $\hat{n}$ case. Ion guiding-center quantities are shown on the left, while electron quantities are shown on the right. The density and $\beta$ plots have been normalized to the source particle rate scaling factor $\hat{n}$, so that the shape and gradient scale lengths of the profiles can be easily compared as $\beta \sim \hat{n}$ increases. The source region is indicated in gray with black dotted lines denoting the FWHM of the source profile. The profiles are quite similar across all cases, which indicates that electromagnetic effects are not playing a significant role in setting the midplane profiles.

We should note that experimental SOL profiles on NSTX are much steeper, falling off to near zero within a few centimeters of the last-closed-flux-surface. There are many effects that we are not currently modeling that could reduce transport and make the profiles steeper, including using the magnetic geometry from the experiment with magnetic shear and an X-point. These effects will be included in future work, but for now, we do not expect agreement between our profiles and the experiment. Nonetheless, we can still investigate interesting physical aspects of the simulations and the influence of electromagnetic effects on the dynamics. Furthermore, strong magnetic shear could make electromagnetic dynamics more relevant at experimental values of heating power and $\beta$.

Even though we cannot currently reproduce the steep gradients seen in experiments, by increasing $\beta \sim \hat{n}$, we can reproduce experimentally relevant values of the MHD ballooning parameter $\alpha$. In our helical geometry, $\alpha$ should be defined as $\alpha = \gamma_{int}^2/(k_\parallel^2 v_A^2) = L_z^2 \beta/(\pi^2 R L_p)$, with $v_A$ being the Alfvén speed, $\gamma_{int} = \sqrt{2} c_s/\sqrt{R L_p}$ the ideal interchange growth rate, and $k_\parallel = \pi/L_z$ the most unstable parallel wavelength for a helical system. [In the standard circular tokamak approximation, one takes $k_\parallel \sim 1/(qR)$ or $L_z \sim \pi qR$ to obtain the standard definition of $\alpha = q^2 R \partial \beta/\partial r = q^2 R \beta/L_p$.] This gives an ideal ballooning limit at $\alpha = 1$ for helical geometry with no magnetic shear. In Fig. 4, we show midplane profiles of $\alpha$ for each case. The range of maximum values $\alpha = 0.3 - 1.5$ is similar to the range observed in experiments, although in realistic geometry the ideal limit can be somewhat higher so that experimental $\alpha$ values may not exceed the ballooning limit.[24,25,63] The fact that we observe midplane values of $\alpha$ that exceed the ideal ballooning limit $\alpha = 1$ for our geometry will be examined in a separate paper.[64]

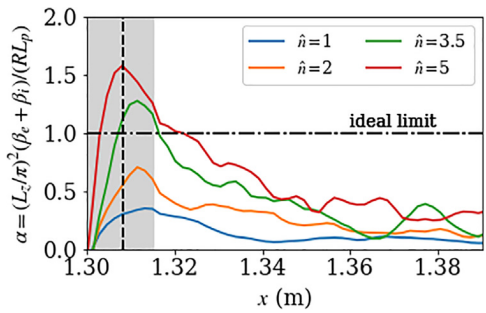

**FIG. 4.** Radial midplane profiles of $\alpha = L_z^2 \beta/(\pi^2 R L_p)$, the MHD ballooning stability parameter in helical geometry with no magnetic shear. The ideal ballooning limit $\alpha = 1$ is shown as a dashed–dotted horizontal line.

A critical issue for future tokamak experiments and reactors is the heat exhaust problem, with large heat loads posing a risk to the survivability of the divertor plates. The heat-flux width is determined by the competition between parallel and perpendicular (cross field) transport in the SOL. Thus, it is important to develop high-fidelity turbulence modeling capability to be able to predict the heat loads and heat-flux widths on the divertor plates. While our present simulations do not have the realistic X-point geometry (including both closed- and open-field-line regions) or neutral particle dynamics required to produce experimentally relevant heat flux predictions, we can still examine the heat flux profiles that result from our simulations. For each species $s$, we can compute the heat flux profile to the lower endplate at $z = -L_z/2$ via

$$Q_{\parallel s}^{\text{end}} = \left\langle \int H_s \mathcal{J} f_s \dot{\mathbf{R}} \cdot \hat{\mathbf{b}} \, d^3 \boldsymbol{v}|_{z=-L_z/2} \right\rangle, \quad (18)$$

with the brackets $\langle \ldots \rangle$ denoting an average in $y$ and time. Here, we include the potential energy via the Hamiltonian, $H_s = m_s v^2/2 + q_s \Phi$, to account for slowing of electrons and acceleration of ions due to the implied potential drop from the sheath entrance (which is the boundary of the simulation) to the grounded wall.

We plot the radial profiles of this quantity normalized to $\hat{n}$ for each case in Fig. 5, with the ion heat flux on the left and the electron heat flux on the right. There are no significant differences in the ion heat flux profiles across the scan, but there is a noticeable trend in the electron heat flux profiles. As $\hat{n}$ increases, the electron heat flux profile broadens and the peak decreases, with the peak flux about 25% lower in the $\hat{n} = 5$ case than the base ($\hat{n} = 1$) case. Differences in the heat flux profiles between electrons and ions reflect differences in the competition between parallel and perpendicular transport, with electrons flowing much more quickly to the endplates so that there is less perpendicular spreading of heat. While the ion heat flux is larger than the electron heat flux in these cases, the electron heat flux dominates in some present experiments and is expected to dominate in ITER.[6] This makes it important to understand turbulent broadening mechanisms for the electron heat flux. Thus, we will primarily focus on the electron heat flux in the following analyses.

Note that while the heat-flux width and the other parameters are influenced by the width of the *ad hoc* source itself, the shape of the source is identical for all cases, as shown by the black dotted lines indicating the source FWHM. This means that the (relative) differences in the profile widths and heights due to broadening effects are physical. Nonetheless, since the absolute peak values and widths are sensitive to the source parameters, a comparison to experimental divertor fluxes is out of the scope of this work; this would likely require the inclusion of closed-field-line regions to eliminate the need for *ad hoc* sourcing of the SOL, since much of the sourcing of heat (and particles, depending on where ionization occurs) for the SOL comes from the core in tokamaks.

Following Ref. 65, we fit the electron parallel heat flux profiles to the function

$$q(\bar{x}) = \frac{q_0}{2} \exp\left[\left(\frac{S}{2\lambda_q}\right)^2 - \frac{\bar{x}}{\lambda_q}\right] \text{erfc}\left(\frac{S}{2\lambda_q} - \frac{\bar{x}}{S}\right),$$
$$\bar{x} = x - x_0, \quad (19)$$

with free constant parameters $S$, $\lambda_q$, $q_0$, and $x_0$. This function is the convolution of an exponentially decaying profile (with decay length $\lambda_q$) and a Gaussian function (with width $S$), which models competition between perpendicular and parallel heat transport. We constrain the fit only within the source region $1.285 \leq x \leq 1.315$ m (shaded).

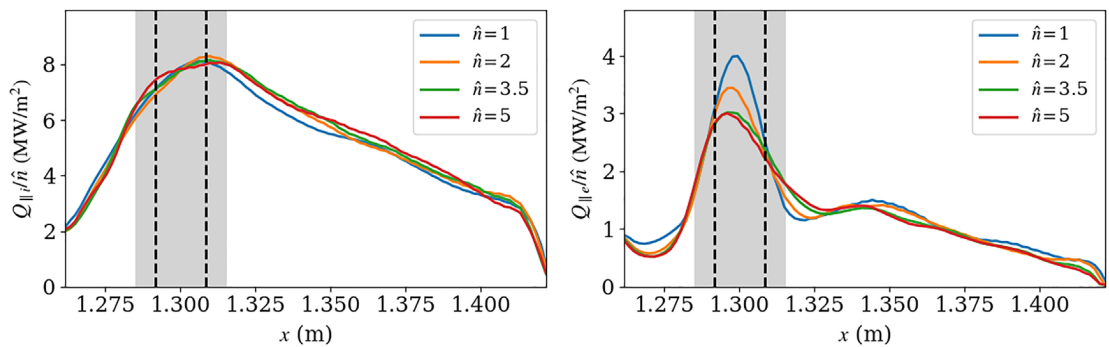

**FIG. 5.** Parallel heat flux to the lower endplate for ions (left) and electrons (right), computed via (18), normalized to the density scaling factor $\hat{n}$. There is a noticeable broadening of the electron heat flux profiles as $\hat{n}$ increases with the peak flux about 25% lower in the $\hat{n} = 3.5$ and $\hat{n} = 5$ cases than the base ($\hat{n} = 1$) case. There is a little change in the ion heat flux profiles.

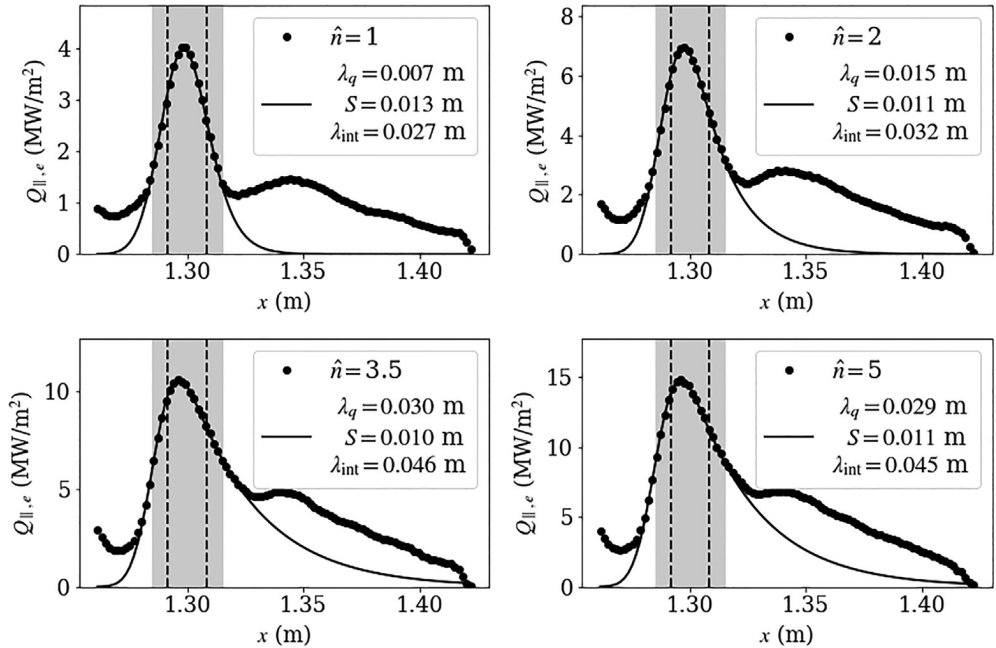

**FIG. 6.** Radial profiles of the electron parallel heat flux to the lower endplate for each case with the peaks fitted to (19). The heat-flux width $\lambda_q$ increases by more than a factor of 4 between the $\hat{n} = 1$ and $\hat{n} = 3.5$ cases, which gives a quantitative measure of the turbulent broadening due to electromagnetic effects.

The fit is shown by the solid lines in Fig. 6, with the fitting parameters $\lambda_q$ and $S$ explicitly shown. We also show the integral power decay length,[66] defined as $\lambda_{int} = \int q(x)dx/Q_{max}$ with $Q_{max}$ taken to be the maximum value of $Q_{\parallel,e}$ and the integral taken over the entire radial domain. This parameter allows the peak heat load to the divertor target to be related to the total power deposited on the divertor target, which is important for power handling in experiments.[65] The heat-flux width $\lambda_q$ increases by more than a factor of 4 between the $\hat{n} = 1$ ($\lambda_q = 0.7$ cm) and $\hat{n} = 3.5$ ($\lambda_q = 3$ cm) cases, which gives a quantitative measure of the turbulent broadening due to electromagnetic effects. The spreading parameter $S \approx 1$ cm is roughly constant for all cases, and the integrated decay length $\lambda_{int}$ increases by about 70% between the $\hat{n} = 1$ and $\hat{n} = 3.5$ cases. Despite larger relative changes in $\lambda_q$ and $\lambda_{int}$, the peak heat flux only drops by about 25%, in part because the lower $\hat{n}$ cases have a larger fraction of the power in the far SOL that is not captured by the functional form of (19).

The broadening of the electron heat-flux width is an indication of increased upstream cross field (perpendicular) transport in the high $\hat{n}$ cases. We confirm this by computing the perpendicular heat flux at the midplane, defined as

$$Q_{\perp e} = \langle \tilde{p}_e \tilde{v}_r \rangle + \langle \tilde{q}_{\parallel e} \tilde{b}_r \rangle. \quad (20)$$

Here, the first term is the contribution from the $E \times B$ drift, with $v_r = E_r/B = -(1/B)\partial\Phi/\partial y$ and $p_e$ being the electron pressure, and the second term is the flux due to magnetic flutter, with $b_r = (1/B)\partial A_\parallel/\partial y$ and $q_{\parallel e}$ being the electron parallel heat flux. The tilde indicates the fluctuation of a time-varying quantity, defined as $\tilde{F} = F - \bar{F}$ with $\bar{F}$ being the time average of $F$. The brackets $\langle F \rangle$ denote an average in $y$ and time. Indeed, in Fig. 7, we see that there is about a 10% increase in the perpendicular electron heat flux near the edge of the source region as $\hat{n}$ increases, relative to the input SOL power (since we have normalized by $\hat{n}$). This is still a relatively small increase in the transport, which may contribute to the fact that there was little change with increasing $\hat{n}$ in the midplane radial profiles in Fig. 3.

To further investigate the competition between parallel and perpendicular transport, in Fig. 8 we show the difference in normalized parallel (a) and perpendicular (b) heat flux between $\hat{n}=1$ and $\hat{n}=5$ cases. In (a), red regions indicate $\frac{1}{5}|Q_{\|e,\hat{n}=5}| > |Q_{\|e,\hat{n}=1}|$, so that the parallel heat flux (which is always directed toward the endplates, hence the absolute value) is relatively stronger in the $\hat{n}=5$ case; blue regions indicate the opposite, where the parallel heat flux is relatively weaker in the $\hat{n}=5$ case. Similarly in (b), red regions indicate $\frac{1}{5}Q_{\perp e,\hat{n}=5} > Q_{\perp e,\hat{n}=1}$, so that the perpendicular heat flux is relatively stronger in the $\hat{n}=5$ case, while blue regions indicate that the perpendicular heat flux is relatively weaker in the $\hat{n}=5$ case. In the vicinity of the source region (which extends from $-2 < z < 2$ m), we can see a clear trend that to the right of the source peak ($x > 1.3$ m), the perpendicular heat flux is relatively stronger in the $\hat{n}=5$ case (large red region in b). Conversely of the source peak ($x < 1.3$ m), the perpendicular heat flux is relatively weaker (blue band). This is consistent with the trend we saw near the midplane in Fig. 7. Further downstream, for $|z| > 2$ m (where there is no source), the relative difference in perpendicular transport is smaller and actually changes sign, so that there is slightly less perpendicular transport near the endplates in the $\hat{n}=5$ case relative to the $\hat{n}=1$ case. The dominant change past $|z|=2$ m is in the parallel heat flux (a). For $|z| > 2$ m, the parallel heat flux is relatively weaker near $x=1.3$ m in the $\hat{n}=5$ case (blue bands), while it is stronger outside the source FWHM (red sidebands). This produces the broader parallel heat flux profile seem in the $\hat{n}=5$ case in Fig. 5.

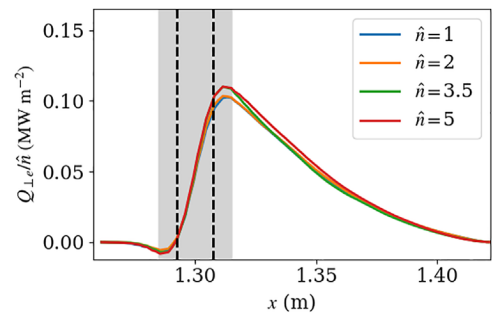

**FIG. 7.** Radial profile of the cross field (perpendicular) electron heat flux near the midplane ($z = 0$), computed via (20), and normalized to $\hat{n}$. There is about a 10% increase in heat flux near the edge of the source region (gray) as $\hat{n}$ increases. This is still a relatively small increase in the transport, which may contribute to the fact that there was a little change with increasing $\hat{n}$ in the midplane radial profiles in Fig. 3. Nonetheless, this small increase in the perpendicular transport appears to be enough to produce the broadening of the electron heat flux to the endplates seen in Fig. 5.

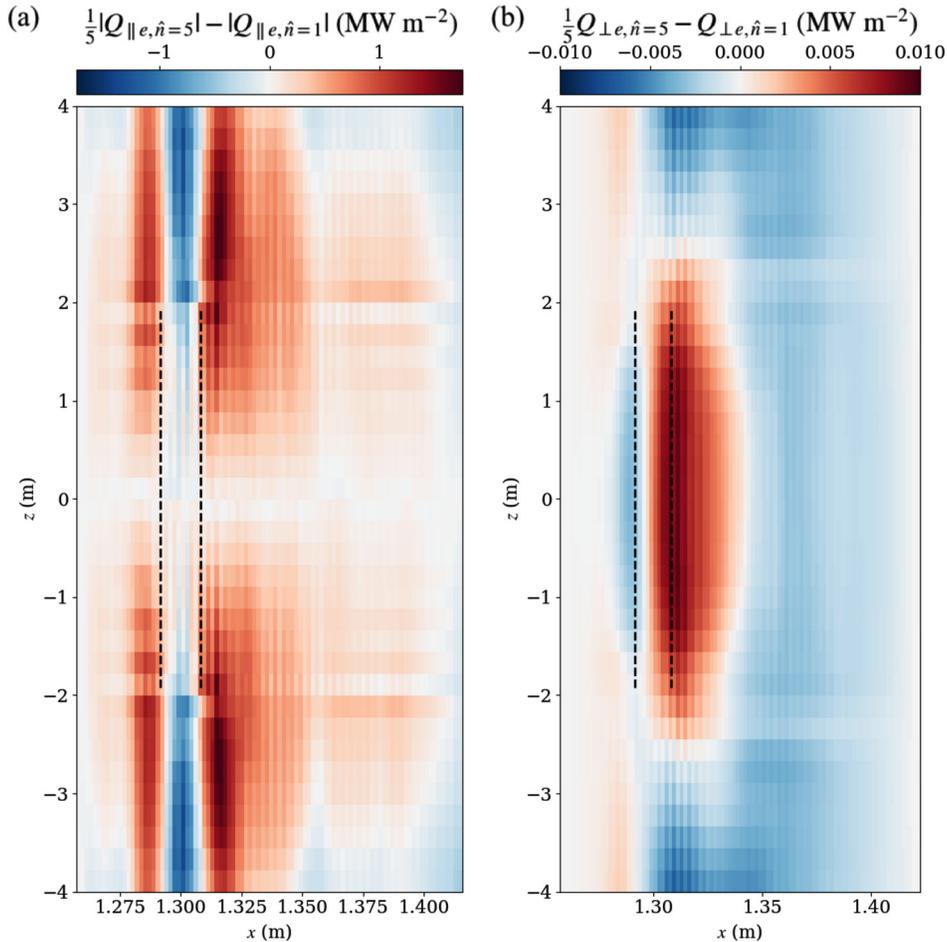

**FIG. 8.** Difference in normalized parallel (a) and perpendicular (b) heat flux between $\hat{n}=1$ and $\hat{n}=5$ cases plotted in the x–z plane (averaged over y and time). In (a), red regions indicate $\frac{1}{5}|Q_{\|e,\hat{n}=5}| > |Q_{\|e,\hat{n}=1}|$, so that the parallel heat flux is relatively stronger in the $\hat{n}=5$ case; blue regions indicate the opposite, where the parallel heat flux is relatively weaker in the $\hat{n}=5$ case. Similarly, in (b), red regions indicate $\frac{1}{5}Q_{\perp e,\hat{n}=5} > Q_{\perp e,\hat{n}=1}$, so that the perpendicular heat flux is relatively stronger in the $\hat{n}=5$ case, while blue regions indicate that the perpendicular heat flux is relatively weaker in the $\hat{n}=5$ case; the strongest red region represents about a 10% relative increase in perpendicular transport in the $\hat{n}=5$ case, consistent with the peak values in Fig. 7. Black dotted lines denote the radial FWHM of the source profile, which extends from $-2 < z < 2$ m.

These plots show that the turbulent widening of the heat flux profile due to electromagnetic effects is mostly happening in the vicinity of the source region, and that the widening essentially stops after contact with the source is lost. The two-point model for diverted tokamaks usually takes the upstream location to be the divertor entrance (near the X point), since that is where the power loading of the SOL cuts off.[67] Our results are complementary to this picture, with the extension that turbulent broadening also cuts off when contact with the source is lost. As a result, the parallel heat flux $Q_\|$ is relatively constant downstream of the source region (or X point), consistent with the assumptions of the two-point model.

We would also like to contextualize our results within the context of Goldston's heuristic drift (HD) model.[4] A key ingredient of the HD model is the radial magnetic drift, which is absent in our simulations due to the simplified geometry. Nonetheless the heat flux profiles that we obtain are wider than those that would be predicted by the HD model for NSTX. This is due to the fact that the turbulent transport is stronger than drift in our cases. To see this, we can take the magnetic drift velocity as $v_M \sim c_s \rho/R$ and the turbulent $E \times B$ drift velocity as $v_E \sim \gamma_{\text{int}}/k_\perp$. This gives $v_M/v_E \sim k_\perp \rho \sqrt{L_p/R} \ll 1$, since $R/L_p \gg 1$ and $k_\perp \rho \lesssim 1$. We do not try to justify this for realistic geometry, where turbulent transport is expected to be weaker due to strong magnetic shear in the SOL and other stabilizing effects.

## V. CONCLUSIONS

In this paper, we performed a study of the effects of increasing $\beta$ on scrape-off layer turbulence dynamics using electromagnetic gyrokinetic simulations of a helical scrape-off layer model with NSTX-like parameters. By increasing the source particle rate and thus $\beta$, we reached experimentally relevant values of the MHD ballooning parameter $\alpha = 0.3 - 1.5$. The main result was a moderate broadening of the electron heat-flux width to the endplates as $\beta$ (and $\alpha$) increased, with over a fourfold increase in $\lambda_q$, a twofold increase in $\lambda_{\text{int}}$, and a 25% decrease in the peak electron heat flux to the endplates. We attribute this to a slight 10% increase in cross field transport at higher $\beta$ (and $\alpha$) with the turbulent broadening mostly occurring in the vicinity of the upstream source region. A secondary result was the fact that despite the increase in cross field transport, higher $\beta$ did not significantly affect the midplane profiles shapes and gradient scale lengths of density or pressure.

Another key result is that the GKEYLL code is able to robustly handle relatively large magnetic fluctuations in a parameter regime with $\alpha$ comparable to the ideal ballooning limit. Modifications of the ballooning limit due to sheath effects and finite radial eigenmode structure will be reported in a separate paper.[64] Since our simulations used a simplified helical model of the tokamak SOL with no magnetic shear, future work will study electromagnetic effects in the SOL in more realistic geometry. In a real experiment, one might expect steeper pressure gradients, stronger magnetic fields, longer connection lengths, and magnetic shear, all of which could push the system into a more electromagnetic regime at experimental $\beta$ levels. Including these effects in future simulations will allow closer comparison with experimental SOL measurements.

## ACKNOWLEDGMENTS

We would like to thank Tess Bernard, Petr Cagas, James Juno, and other members of the GKEYLL team for helpful discussions and support, including the development of the postgkyl post-processing tool, which facilitated the creation of many figures in this paper. We acknowledge additional helpful discussions with Matt Kunz, Walter Guttenfelder, Bill Dorland, and Federico Halpern. Research support came from the U.S. Department of Energy: N.R.M. was supported by the DOE Fusion Energy Sciences Postdoctoral Research Program administered by the Oak Ridge Institute for Science and Education (ORISE) for the DOE via Oak Ridge Associated Universities (ORAU) under DOE Contract No. DE-SC0014664; G.W.H., A.H., and M.F. were supported by the Partnership for Multiscale Gyrokinetic Turbulence (MGK) and the High-Fidelity Boundary Plasma Simulation (HBPS) projects, part of the U.S. Department of Energy (DOE) Scientific Discovery Through Advanced Computing (SciDAC) program, via DOE Contract No. DE-AC02–09CH11466 for the Princeton Plasma Physics Laboratory. Computations were performed on the Stellar cluster at Princeton University and the Cori cluster at the National Energy Research Scientific Computing Center (NERSC). All opinions expressed in this paper are the authors' and do not necessarily reflect the policies and views of DOE, ORAU, or ORISE.

## AUTHOR DECLARATIONS
### Conflict of Interest

The authors have no conflicts to disclose.

## DATA AVAILABILITY

The data that support the findings of this study are available from the corresponding author upon reasonable request.

## APPENDIX A: PROBLEMS WITH UNDER-RESOLVED HIGH-BETA SIMULATIONS

The simulations presented in the body of this paper used $48 \times 96$ cells in the perpendicular dimensions. Initially, these simulations were performed with $16 \times 32$ cells in the perpendicular dimensions, three times less resolution in each of the perpendicular dimensions. Recall that since we are using a piecewise-linear ($p=1$) discontinuous Galerkin discretization scheme, one should double the number of cells in each dimension to obtain the total number of degrees of freedom. Figure 9 shows the midplane electron beta profile from the lower-resolution ($16 \times 32$) simulations of the $\hat{n} = 1$ (left) and $\hat{n} = 5$ (right) cases, overlaid with profiles from the corresponding higher-resolution simulations ($48 \times 96$) presented in the body of this paper. While there is a little difference in the $\hat{n} = 1$ profile as the resolution is increased, there is a substantial difference in the profile in the low-resolution $\hat{n} = 5$ case. We have also included an intermediate resolution $\hat{n} = 5$ case with $32 \times 64$ cells to verify that the high-resolution $\hat{n} = 5$ case is numerically resolved in the perpendicular dimensions. It is beyond the scope of this work to investigate in detail the small-scale dynamics that causes this difference, but some clues can be found by examining the structure of the parallel current. In Fig. 10, we plot a perpendicular cut of the parallel current $j_\parallel$, taken at $z = 2$ m (halfway between the midplane and the upper endplate). In the high resolution cases ($48 \times 96$, right), the current structures are well-resolved, with fine-scale structure and current sheets visible. Meanwhile, in the low-resolution cases ($16 \times 32$, left) the scale of the structures is limited to the grid scale. This seems to modify the dynamics, but only at high beta where the currents are more intense.

## APPENDIX B: GETTING GKEYLL AND REPRODUCING RESULTS

Readers may reproduce our results and also use GKEYLL for their applications. The code and input files used here are available online. Full installation instructions for GKEYLL (gkyl.readthedocs.io) are provided on the GKEYLL website (gkyl.readthedocs.io). The code can be installed on Unix-like operating systems (including Mac OS and Windows using the Windows Subsystem for Linux) either by installing the pre-built binaries using the conda package manager (https://www.anaconda.com) or building the code via sources. The input files used here are under version control and can be obtained from the repository at https://github.com/ammarhakim/gkyl-paper-inp/tree/master/2022_PoP_EMGK_broadening.

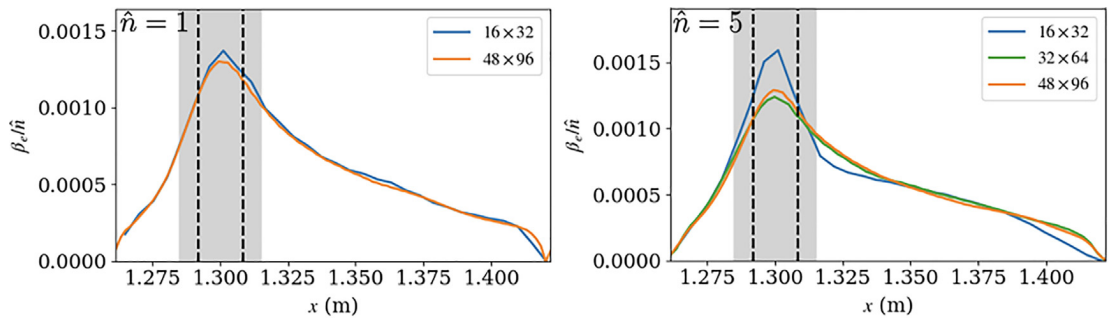

FIG. 9. Comparison of electron $\beta$ profiles with low perpendicular resolution (16 × 32, blue) and high resolution (48 × 96, orange) for the $\hat{n}=1$ (left) and $\hat{n}=5$ (right) cases. While there is a little difference in the $\hat{n}=1$ profile as the resolution is increased, there is a substantial difference in the profile in the low-resolution $\hat{n}=5$ case. Also included is an intermediate-resolution $\hat{n}=5$ case (32 × 64, green) to verify numerical convergence.

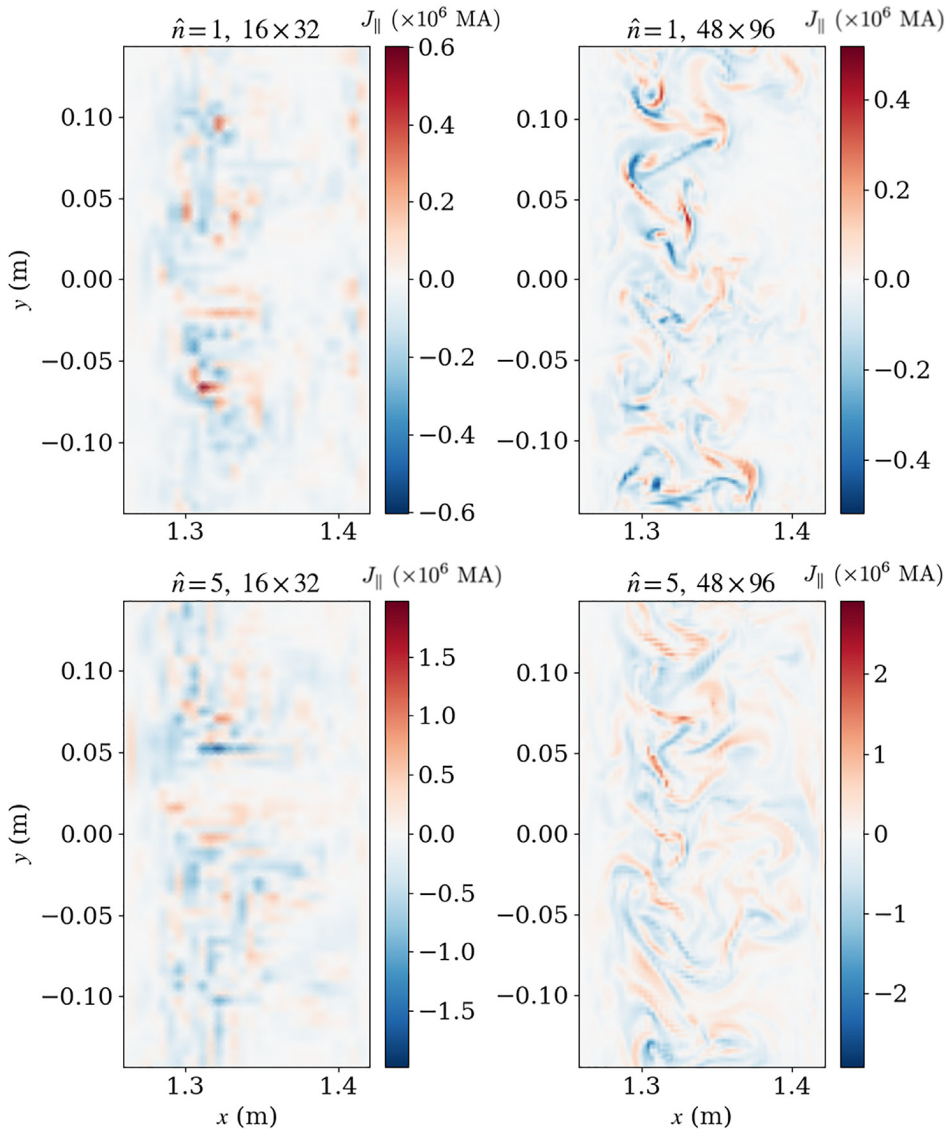

FIG. 10. Perpendicular cuts of the parallel current $j_\parallel$, taken at $z=2$ m (halfway between the midplane and the upper end-plate). In the high perpendicular resolution cases (48 × 96, right), the current structures are well-resolved with fine-scale structure and current sheets visible. Meanwhile, in the low-resolution cases (16 × 32, left), the scale of the structures is limited to the grid scale. In the high beta ($\hat{n}=5$, bottom) case, the currents are more intense, and improperly resolving the current dynamics could be related to the differences in $\hat{n}=5$ profiles seen in Fig. 9.